\documentclass[aps,prb,floatfix,twocolumn,reprint,amsmath,amssymb,superscriptaddress]{revtex4-2}
\usepackage{amsfonts}
\usepackage{mathrsfs}
\usepackage{amsmath}
\usepackage{color}
\usepackage{natbib}
\usepackage{graphicx}
\usepackage{bm}
\usepackage{amssymb}
\usepackage{xspace}
\usepackage{epstopdf}
\usepackage{dcolumn}
\usepackage{multirow}
\usepackage[colorlinks=true, letterpaper=true, pdfstartview=FitV, linkcolor=blue, citecolor=blue, urlcolor=blue]{hyperref}
\usepackage{wrapfig}
\usepackage{diagbox}

\makeatletter

\newcommand{\Rmnum}[1]{\expandafter\@slowromancap\romannumeral #1@}
\makeatother

\begin{document}
\title{Giant Magnetostriction by Design: A First-Principles Screening of Co-based Heusler Alloys}

\author{Pengju Wu}
\author{Jie Du}
\affiliation{School of Material Science and Engineering, Tiangong University, Tianjin 300387, China}
\author{Liang Yao}
\author{Hang Li}
\affiliation{School of Electronic and Information Engineering, Tiangong University, Tianjin 300387, China}
\author{Xiaodong Zhou}
\affiliation{School of Physical Science and Technology, Tiangong University, Tianjin 300387, China}
\author{Tao Zhu}\thanks{Corresponding authors}\email{zhutao@tiangong.edu.cn}
\author{Wenhong Wang}\thanks{Corresponding authors}\email{wenhongwang@tiangong.edu.cn}
\affiliation{School of Electronic and Information Engineering, Tiangong University, Tianjin 300387, China}

\begin{abstract}
	The pursuit of high-performance, rare-earth-free magnetostrictive materials is crucial for advancing technologies in sensing, actuation, and microelectromechanical systems. Heusler alloys represent a promising, yet underexplored, class of materials for this purpose. In this work, we perform a systematic first-principles investigation of the magnetostrictive properties of 25 Co-based full Heusler alloys, Co$_2$YZ (Y = V, Cr, Mn, Fe, Co; Z = Al, Ga, Si, Ge, Sn). Our screening identifies 10 compounds with large predicted magnetostriction ($|\lambda_{001}| > 100$~ppm), highlighted by Co$_3$Si with a giant value of -966~ppm. Furthermore, we demonstrate two effective strategies for engineering magnetostriction: (i) tuning the Fermi level, which enhances the magnetostriction of Co$_3$Sn to -905~ppm via Sb doping, and (ii) amplifying the spin-orbit coupling, which boosts the magnetostriction of Co$_2$CrGa to a colossal -1008~ppm through Re substitution. Our analysis reveals a general predictive rule, uncovering a linear relationship between the magnetostriction and the choice of the Y-site transition metal. This work not only identifies novel candidates for magnetostrictive applications but also establishes clear, physically-grounded design principles to accelerate the discovery of new functional magnetic materials.
\end{abstract}
\maketitle
\section{Introduction}
Magnetostriction is the phenomenon wherein a material's dimensions change in response to an applied magnetic field. This property is fundamental to a wide range of technologies, including sensors, actuators, transducers, and microelectromechanical systems (MEMS) \cite{add1,add2,add3,add4}. Since its initial discovery in iron by J. P. Joule in 1842 \cite{add5}, the pursuit of materials exhibiting large magnetostriction has been a significant research endeavor. This search has progressed from elemental iron ($\sim$70 ppm) \cite{add6,add7} and Fe-based alloys like FeGa ($\sim$400 ppm) \cite{add8,add9,add10} to rare-earth--transition-metal compounds such as Terfenol-D (Tb$_{0.3}$Dy$_{0.7}$Fe$_2$), which displays a giant magnetostriction of up to 2000 ppm \cite{add11,add12}. Materials with magnetostriction exceeding 100 ppm are generally classified as having large magnetostriction \cite{add12A}. However, the preeminence of Terfenol-D is challenged by practical drawbacks, including its inherent brittleness, the high magnetic fields required for saturation, and significant material costs, which collectively limit its broader application \cite{add13,add14,add15}. 

In the search for alternative high-performance magnetostrictive materials, Heusler alloys have emerged as a promising class of compounds~\cite{heusler}. These ternary intermetallics typically adopt the L2$_{1}$ crystal structure with the stoichiometric formula X$_2$YZ (space group Fm$\bar{3}$m), where X and Y are transition metals and Z is a main-group element \cite{add16,add17}. The Heusler family is renowned for a diverse range of functional properties—including half-metallicity \cite{add18}, superconductivity \cite{add19}, and the anomalous Nernst effect \cite{add20,add21}—often rooted in their tunable electronic and magnetic structures. The potential for significant magnetostriction within this family is exemplified by Ni$_2$MnGa, which exhibits a large magnetostrictive response (from -130 ppm to -250 ppm) associated with its martensitic phase transformation \cite{add22,add23}. Despite this potential, and in contrast to the extensive research on Fe-based systems, a systematic exploration of magnetostriction across the broader Heusler family remains limited, with many compounds yet to be investigated both experimentally and theoretically. 

In this paper, we present a systematic theoretical investigation of the magnetostrictive properties of 25 Co-based full Heusler alloys, Co$_2$YZ (Y = V, Cr, Mn, Fe, Co; Z = Al, Ga, Si, Ge, Sn), using first-principles calculations based on density functional theory (DFT). Our calculations reveal that 10 of these compounds are predicted to exhibit large magnetostriction, with values exceeding 100 ppm. Furthermore, we explore strategies to enhance the magnetostrictive performance, such as tuning the electronic density of states near the Fermi level and incorporating elements with strong spin-orbit coupling (SOC). This work not only identifies new potential candidates for large magnetostrictive materials but also provides theoretical guidance for the future design and discovery of functional Heusler alloys, thereby broadening their potential applications.

\section{Computational Details}

\begin{figure}[ht!]
	\centering
	\includegraphics[width=8.5cm]{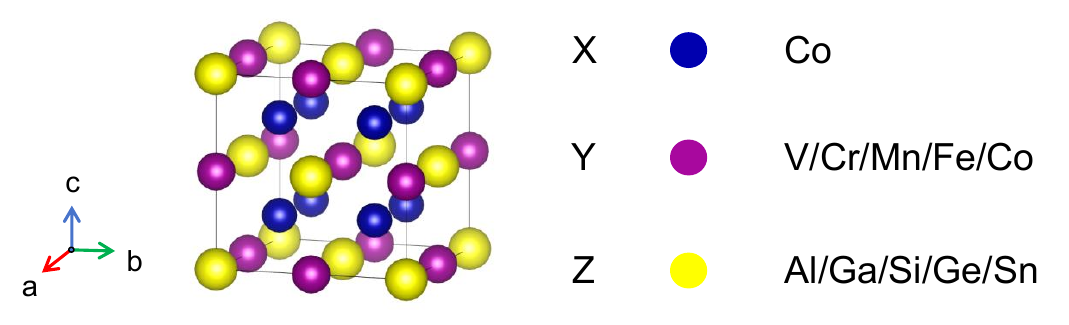}
	\caption{The L2$_1$ crystal structure of a full Heusler alloy Co$_2$YZ. Blue, purple, and yellow spheres represent the Co, Y (V, Cr, Mn, Fe, Co), and Z (Al, Ga, Si, Ge, Sn) atoms, respectively.}
\label{fig1}
\end{figure}

All first-principles calculations were performed using the Vienna Ab initio Simulation Package (VASP) \cite{add24,add25}, which is based on density functional theory (DFT). The projector augmented-wave (PAW) method was employed to describe the electron-ion interactions \cite{add26}. For the exchange-correlation functional, we utilized the generalized gradient approximation (GGA) as parameterized by Perdew, Burke, and Ernzerhof (PBE) \cite{add27,add28}.

The crystal structures of the Co-based full Heusler alloys were fully relaxed to determine their equilibrium lattice parameters and atomic positions. Since the primitive Heusler alloy unit cell is not cubic, all magnetostriction calculations were conducted using the conventional cubic L2$_1$ cell structure with equal lattice parameters ($a=b=c$). For structural optimization, the Brillouin zone was sampled using a 6$\times$6$\times$6 $\Gamma$-centered $k$-point mesh \cite{add29}. A plane-wave energy cutoff of 500 eV was applied. The structural relaxation was continued until the residual forces on each atom were below 0.01 eV/\AA, and the total energy converged to within $10^{-6}$ eV between successive self-consistent field cycles.

As the standard PAW potentials include scalar-relativistic effects, the magnetocrystalline anisotropy energy (E$_{\text{MCA}}$) was calculated by including SOC in a second, non-collinear variational step \cite{add30,add31}. To ensure sufficient numerical precision for the small energy differences involved in MCA, we increased the plane-wave cutoff to 560 eV and the $k$-point mesh was increased to 12$\times$12$\times$12 combined with a stricter total energy convergence criterion of $10^{-8}$ eV. E$_{\text{MCA}}$ was determined using the total energy difference method, defined as the energy difference between magnetization orientations along two orthogonal crystallographic directions \cite{add32}:
\begin{equation}
	E_{\text{MCA}} = E_{[100]} - E_{[001]}
\end{equation}
where $E_{[100]}$ and $E_{[001]}$ are the total energies with the magnetization aligned along the [100] and [001] axes, respectively.

\section{Results and Discussion}

\begin{table}[ht!]
	\centering
	\caption{Calculated tetragonal magnetostriction coefficients ($\lambda_{001}$) in parts per million (ppm) for the 25 Co-based full Heusler alloys, Co$_2$YZ. The Y-site element defines the rows, and the Z-site element defines the columns. Values are predicted from first-principles calculations. Ten compounds are predicted to exhibit large magnetostriction, defined as $|\lambda_{001}| >$ 100 ppm, and are highlighted in bold.}
	\label{tab1}
	\begin{tabular}{|c|m{1cm}|m{1cm}|m{1cm}|m{1cm}|m{1cm}|} \hline
		\diagbox{Y-site}{Z-site} & Al & Ga & Si & Ge & Sn \\ \hline
		V  & 36 & 57 & \textbf{105} & \textbf{156} & \textbf{113}  \\ \hline
		Cr  & \textbf{112} & \textbf{184} & 67 & 53 & 92   \\ \hline
		Mn  & 74 & 79 & 20 & 33 & 32  \\ \hline
		Fe  & 15 & 20 & 36 & 43 & 45   \\ \hline
		Co  & \textbf{-174} & \textbf{-243} & \textbf{-966} & \textbf{-303} & \textbf{-385}	\\ \hline
	\end{tabular}
\end{table}

The tetragonal magnetostriction coefficient, $\lambda_{001}$, can be derived from the strain dependence of the total energy ($E_{\text{tot}}$) and the magnetocrystalline anisotropy energy ($E_{\text{MCA}}$) using the following relation \cite{add12A,add33}:

\begin{equation}
	 \lambda_{001} = \frac{2 dE_{MCA} / d\varepsilon_z}{3 d^2 E_{tot} / d\varepsilon_z^2} = -\frac{b_1}{3c'}.
	\label{eq:lambda001}
\end{equation}
In this expression, $b_1$ is the magnetoelastic coupling coefficient, which quantifies the change in $E_{\text{MCA}}$ with an applied volume-conserving tetragonal strain ($\varepsilon_{z}$). The term $c'$ represents the tetragonal shear modulus, related to the elastic constants $C_{11}$ and $C_{12}$. Equation~\eqref{eq:lambda001} makes it clear that achieving a large magnetostriction requires both a strong magnetoelastic coupling (a large $|b_1|$) and a relatively soft lattice with respect to tetragonal shear (a small $c'$).

To understand the origin of magnetostriction, it is crucial to analyze the factors governing $E_{\text{MCA}}$. Within the framework of second-order perturbation theory, $E_{\text{MCA}}$ arises from the spin-orbit coupling (SOC) and can be expressed as \cite{add35,add36,add37}:
\begin{equation}
	E_{\text{MCA}} \approx \sum_{o,u} \frac{ | \langle \psi_o | \hat{H}_{\text{SOC}} | \psi_u \rangle |^2_{[100]} - | \langle \psi_o | \hat{H}_{\text{SOC}} | \psi_u \rangle |^2_{[001]} }{E_o - E_u}
	\label{eq:emca_pert}
\end{equation}
where $|\psi_o\rangle$ and $|\psi_u\rangle$ are the occupied and unoccupied single-particle states with corresponding energies $E_o$ and $E_u$, and $\hat{H}_{\text{SOC}} = \xi \mathbf{L} \cdot \mathbf{S}$ is the SOC Hamiltonian. This expression highlights two primary routes to engineer a large and strain-sensitive $E_{\text{MCA}}$: (i) incorporating elements with a large SOC strength ($\xi$), and (ii) tailoring the electronic structure to have occupied and unoccupied states, coupled by the angular momentum operator, that are close in energy to the Fermi level (a small energy denominator $E_o - E_u$) \cite{add38}.

The full Heusler alloys investigated in this work, Co$_2$YZ, crystallize in the L2$_1$ structure, as shown in Fig.~\ref{fig1}. In this highly ordered arrangement, Co atoms occupy the Wyckoff sites (0, 0, 0) and (1/2, 1/2, 1/2), while the Y and Z atoms occupy the (1/4, 1/4, 1/4) and (3/4, 3/4, 3/4) sites, respectively. The Co atoms form a rigid framework that governs the hybridization with the Y-site atom. The Y-site transition metal (V, Cr, Mn, Fe, Co) is typically the primary contributor to the total magnetic moment in the system \cite{add39,add40}. The main-group Z-element (Al, Ga, Si, Ge, Sn) serves to modulate the electronic structure, primarily through p-d orbital hybridization, which influences the d-electron count and magnetic exchange interactions \cite{add41,add42}. As shown in Eq.~\ref{eq:emca_pert}, the interplay between the electronic states near the Fermi level and the SOC is therefore the critical factor determining the magnetostrictive performance. This understanding forms the basis of our strategy to enhance magnetostriction by tuning the composition.

To benchmark our computational methodology, we first calculated the magnetostriction of a bcc-based Fe$_7$Ga alloy. Our obtained value of 142 ppm, as shown in Fig.~\ref{fig5} (b) in the Appendix, is in good agreement with previously reported theoretical values~\cite{add43}. This result serves as a useful benchmark for identifying promising candidates within the Heusler family. With this reference, we screened the magnetostrictive properties of the 25 Co$_2$YZ compounds. The results are shown in Table~\ref{tab1} and the general trend of the calculated values is consistent with recent experimental measurement~\cite{heusler}. Moreover, our calculations predict that 10 of these alloys exhibit large magnetostriction ($|\lambda_{001}| > 100$ ppm). Most notably, the compound Co$_3$Si is predicted to exhibit a giant magnetostriction of -966 ppm. This value is exceptionally large for a rare-earth-free material and is comparable in magnitude to that of Terfenol-D. Furthermore, Co$_3$Si with the L2$_1$ structure can be successfully synthesized via arc melting-annealing methodology \cite{add44}, marking it as a highly promising candidate for further investigation.

\begin{figure}[ht!]
	\centering
	\includegraphics[width=8.5cm]{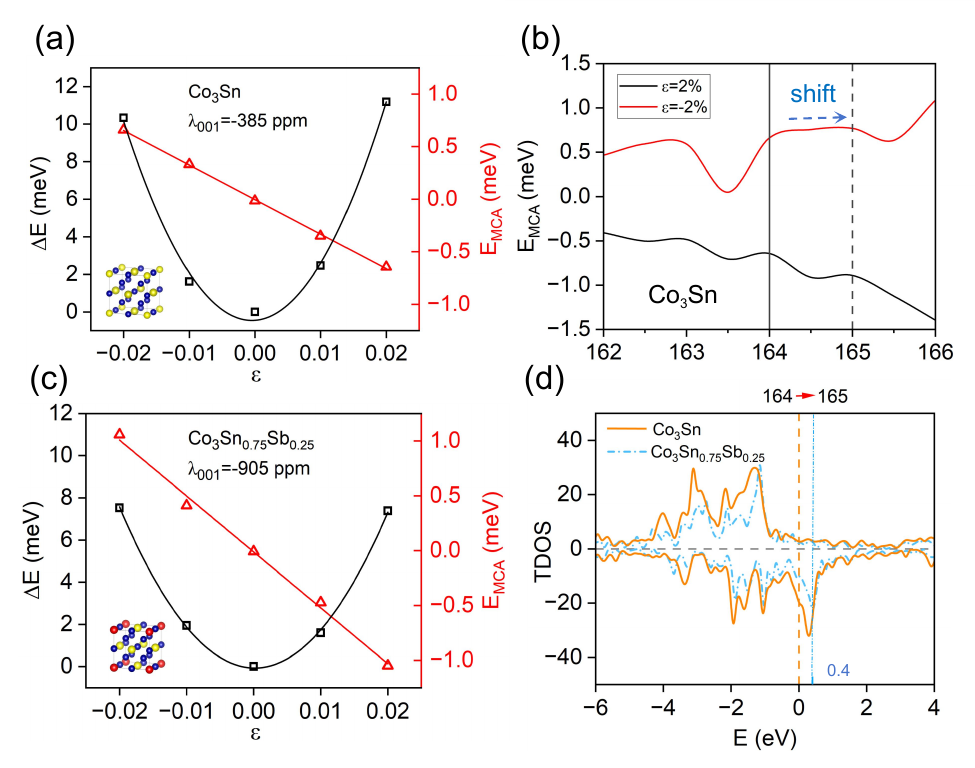}
	\caption{Enhancement of magnetostriction in Co$_3$Sn via electron doping. (a), (c) Total energy ($E_{\text{tot}}$, open squares) and magnetocrystalline anisotropy energy ($E_{\text{MCA}}$, open triangles) as a function of tetragonal strain for pristine Co$_3$Sn and Sb-doped Co$_3$Sn$_{0.75}$Sb$_{0.25}$, respectively. The inset shows the supercell used for the doped calculation. (b) The rigid band model prediction for the dependence of $E_{\text{MCA}}$ on electron count ($N_e$) for Co$_3$Sn under $\pm$2\% strain. (d) The projected density of states (DOS) for Co$_3$Sn (solid line) and Co$_3$Sn$_{0.75}$Sb$_{0.25}$ (dotted-dashed line), The vertical yellow dashed line indicates the Fermi level with Co$_3$Sn as the reference zero. The vertical blue dotted-dashed line represents the shifted Fermi level of Co$_2$Sn$_{0.75}$Sb$_{0.25}$, illustrating the shift of the Fermi level into a DOS peak due to electron doping.}
	\label{fig:2}
\end{figure}

\subsection{Enhancement via Fermi Level Tuning: The Case of Co$_3$Sn}

To demonstrate strategies for enhancing magnetostriction, we first consider Co$_3$Sn, for which our calculations yield a substantial magnetostriction coefficient of $\lambda_{001} = -385$~ppm as shown in Fig.~\ref{fig:2}(a). We explored the potential for further enhancement by manipulating the electronic structure near the Fermi level ($E_F$). Using a rigid band model (RBM), we simulated the effect of electron doping by shifting $E_F$ relative to the fixed band structure of Co$_3$Sn. As shown in Fig.~\ref{fig:2}(b), the RBM predicts an increase in the difference of $E_{\text{MCA}}$ under $\pm$2\% strain for a slight increase in the electron count ($N_e$), suggesting that electron doping is a viable path to higher magnetostriction. We acknowledge that the RBM approximates the electronic band structure as rigid upon doping, with the only effect being a shift in the Fermi level. This approximation neglects effects such as lattice parameter changes, local structural relaxations, and charge redistribution specific to the dopant. Nevertheless, for low doping concentrations, it can serve as a screening guide for higher concentrations. Results in Fig.~\ref{fig:2}(b) identify the direction and approximate magnitude of the required chemical potential shift, which guides the subsequent stoichiometric engineering.

This theoretical prediction can be tested computationally by substituting Sn with an element that provides additional valence electrons, such as Antimony (Sb). We therefore calculated the properties of the doped alloy Co$_3$Sn$_{0.75}$Sb$_{0.25}$, with the crystal structure shown in the inset of Fig.~\ref{fig:2}(c). The results show a remarkable enhancement: the substitution not only increases the magnetoelastic coupling coefficient ($|b_1|$) but also softens the lattice by reducing the tetragonal shear modulus ($c'$). The synergistic effect of these two changes leads to a predicted giant magnetostriction of $\lambda_{001} = -905$~ppm.

The mechanism for this enhancement is revealed by the density of states (DOS), plotted in Fig.~\ref{fig:2}(d). The substitution of Sn with Sb acts similarly to the RBM, shifting the Fermi level upwards with minimal distortion of the overall band structure. This shift moves $E_F$ directly into a sharp peak in the spin-down channel, significantly increasing the density of states at the Fermi level. According to the perturbation theory for $E_{\text{MCA}}$ (Eq.~(\ref{eq:emca_pert})), a higher DOS at $E_F$ can lead to a larger anisotropy by providing more states with small energy denominators ($E_o - E_u$), thus amplifying the effect of SOC and leading to the observed increase in magnetostriction.

\subsection{Enhancement via Spin-Orbit Coupling Engineering: The Case of Co$_2$CrGa}

\begin{figure}[ht!]
	\centering
	\includegraphics[width=8.5cm]{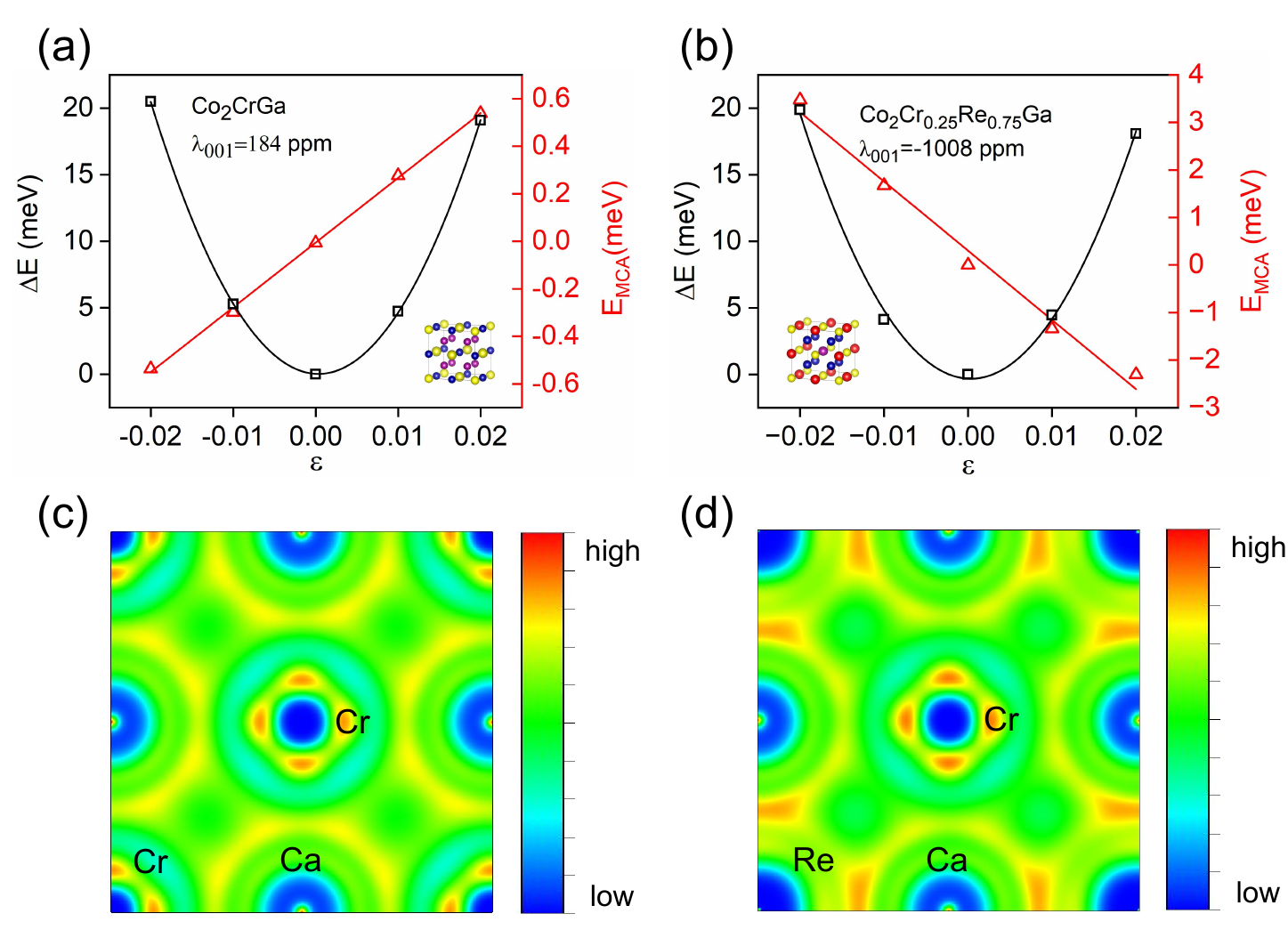}
	\caption{Enhancement of magnetostriction in Co$_2$CrGa via SOC engineering. $E_{\text{tot}}$ (open squares) and $E_{\text{MCA}}$ (open triangles) versus strain for (a) pristine Co$_2$CrGa and (b) Re-doped Co$_2$Cr$_{0.25}$Re$_{0.75}$Ga. Insets show the respective crystal structures. (c) and (d) illustrate the spatial distribution of the electron localization function (ELF) for Co$_2$CrGa and Co$_2$Cr$_{0.25}$Re$_{0.75}$Ga, respectively, within the Cr-Ga atomic plane. Blue and red regions represent low and high degrees of charge localization, respectively.}
	\label{fig:3}
\end{figure}

A complementary strategy for enhancing magnetostriction is to directly increase the SOC strength. To demonstrate this, we start with Co$_2$CrGa, which is predicted to have a magnetostriction of $\lambda_{001} = +184$~ppm as shown in Fig.~\ref{fig:3}(a). We then partially substitute the 3d transition metal Cr with a heavy 5d element, Rhenium (Re), which possesses a much larger intrinsic SOC strength.

The results for the composition Co$_2$Cr$_{0.25}$Re$_{0.75}$Ga are striking, as shown in Fig.~\ref{fig:3}(b). The magnetoelastic coupling coefficient $|b_1|$ is dramatically enhanced, while the lattice stiffness $c'$ remains nearly unchanged. This leads to a predicted magnetostriction of $\lambda_{001} = -1008$~ppm, an order-of-magnitude increase and a value comparable to that of rare-earth-based materials. Notably, the sign of the magnetostriction also reverses, indicating a fundamental change in the strain response of the electronic structure. This sign change suggests a reordering of the key electronic states near the Fermi level that contribute to the anisotropy, with different orbital pairs dominating the perturbation sum in Eq.~(\ref{eq:emca_pert}) under strain \cite{add43}. 

To validate the microscopic origin of the sign change and the giant magnetostriction, we extracted the SOC matrix elements, $\langle d_i | \hat{H}_{soc} | d_j \rangle$, from the self-consistent calculations under 2$\%$ tetragonal strain for magnetization aligned along the [001] and [100] directions. The results show that the substitution of Re dramatically enhances the SOC strength. For example, we found that the coupling magnitude of $\langle d_{x^2-y^2} | \hat{H}_{soc} | d_{xy} \rangle$ increases by approximately two orders of magnitude (from $\sim 0.4$ meV in the pristine sample (Cr) to $\sim 57$ meV in the doped sample (Re)), which explains the ``giant" nature of the magnetostriction in the Re-doped system. Moreover, the anisotropy of the matrix elements for the $d_{xy}$ and $d_{x^2-y^2}$ orbitals undergoes an inversion. For pristine Co$_2$CrGa, the coupling strength is slightly larger for the [001] orientation ($|M_{001}| > |M_{100}|$). In contrast, for Co$_2$Cr$_{0.25}$Re$_{0.75}$Ga, this trend reverses ($|M_{100}| > |M_{001}|$). According to second-order perturbation theory, the MAE is determined by the difference in these squared matrix elements. Consequently, this reversal in the orbital coupling anisotropy directly drives the observed sign change in the magnetostriction coefficient.

On the other hand, the magnetostriction of the system is highly dependent on the electronic states near the Fermi level. To further analyze the impact of Re substitution, we visualized the electron localization function (ELF) before and after doping. As shown in Fig.~\ref{fig:3}(c), the charge in Co$_2$CrGa is largely localized around the Cr sites, exhibiting a delocalized state between Cr and Ga atoms with no significant bonding. In contrast, Fig.~\ref{fig:3}(d) reveals that Re doping leads to increased charge accumulation between Re and Ga atoms, indicative of bond formation. The synergistic effect of this charge redistribution and the strong SOC introduced by Re atoms is responsible for the enhanced magnetostriction in Co$_2$Cr$_{0.25}$Re$_{0.75}$Ga.

\subsection{General Trends and a Predictive Descriptor for Magnetostriction}
\begin{figure*}[ht!]
	\centering
	\includegraphics[width=17cm]{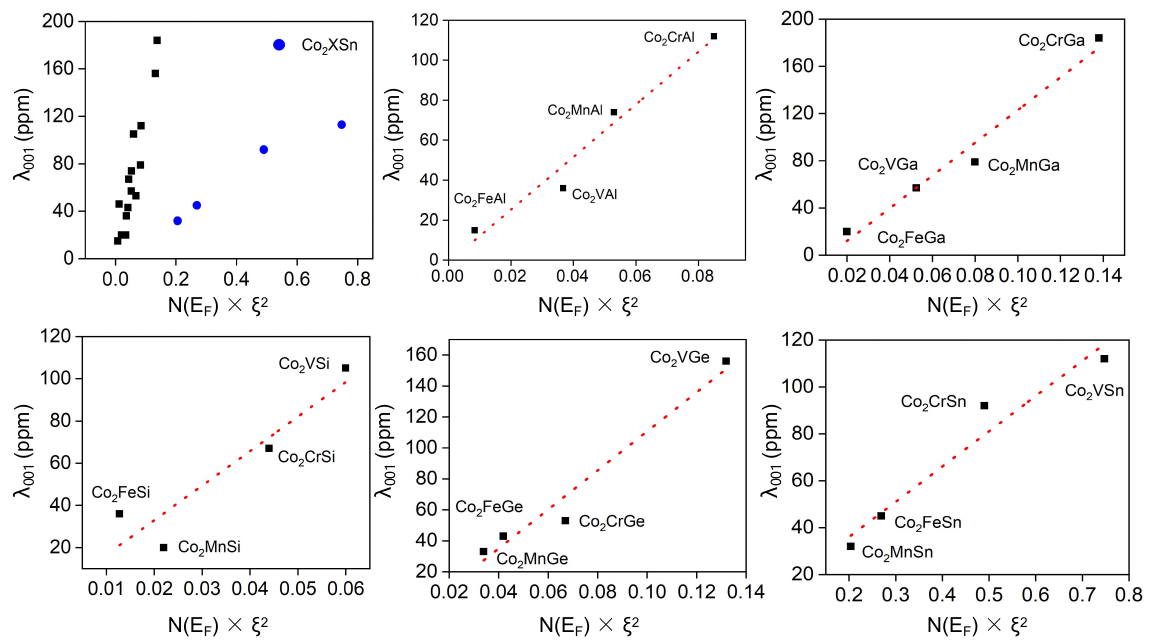}
	\caption{General trends and predictive rules for magnetostriction in Co$_2$YZ (Y = V, Cr, Mn, and Fe) alloys. (a) Correlation between the calculated magnetostriction coefficient ($\lambda_{001}$) and a descriptor combining the density of states at the Fermi level and the square of the Y-site SOC strength, the Co$_2$YSn series is highlighted with blue dots. (b)-(f) Linear relationship observed between $\lambda_{001}$ and the Y-site transition metal for fixed main-group elements Z = Al, Ga, Si, Ge, and Sn, respectively.}
	\label{fig:4}
\end{figure*}
Building on these specific examples, we sought to identify a general trend governing the magnetostriction across the Co$_2$YZ series. As shown in Table~\ref{tab2} in Appendix, the tetragonal shear modulus ($c'$) is relatively stable under element substitution of the Y-site from V to Fe. Based on the theoretical framework (Eqs.~\ref{eq:lambda001} and \ref{eq:emca_pert}), a large magnetostriction is driven by a large magnetoelastic coefficient ($b_1$), which in turn is promoted by strong SOC and a high density of states near the Fermi level. This suggests that a useful descriptor for large magnetostriction might be the product of the SOC strength ($\xi$) and the electronic density of states near $E_F$.

In Fig.~\ref{fig:4}(a), we plot the calculated $\lambda_{001}$ against a descriptor proportional to $N(E_F) \times \xi^2$, where $N(E_F)$ is the density of states at the Fermi level and $\xi$ is the SOC constant of the Y-site atom. The plot reveals a general positive trend, confirming that this simple descriptor successfully captures the key physical ingredients for large magnetostriction in this alloy family.

Furthermore, we observe a remarkable linear trend when the Y-site transition metal is varied for a fixed Z-site main group element, as shown in Figs.~\ref{fig:4}(b)-(f). For instance, in the Co$_2$YSn series (blue dots), $\lambda_{001}$ increases systematically as Y moves from V to Co. This linear behavior is primarily driven by the systematic change in the magnetoelastic coupling coefficient $b_1$, which itself is a consequence of the regular filling of the d-band and the corresponding increase in SOC strength across the 3d series. This finding provides a powerful and intuitive design rule for discovering new Heusler alloys with large magnetostriction.

\section{Summary}

In summary, we have conducted a systematic first-principles investigation into the magnetostrictive properties of 25 Co-based full Heusler alloys, Co$_2$YZ (Y = V, Cr, Mn, Fe, Co; Z = Al, Ga, Si, Ge, Sn). Our primary goal was to explore this vast family of compounds as a potential source of high-performance, rare-earth-free magnetostrictive materials. The screening revealed significant potential, identifying 10 compounds predicted to exhibit large magnetostriction ($|\lambda_{001}| > 100$~ppm). Among these, Co$_3$Si stands out as an exceptionally promising candidate, with a predicted giant magnetostriction of -966~ppm, a value comparable in magnitude to the industry-standard material Terfenol-D.

Beyond identifying individual compounds, we have demonstrated and elucidated two powerful strategies for engineering and enhancing magnetostriction in Heusler alloys. First, by tuning the electronic structure via electron doping—as demonstrated by substituting Sb for Sn in Co$_3$Sn—we showed that aligning the Fermi level with a sharp peak in the density of states can dramatically increase the magnetostriction, achieving a value of -905~ppm. Second, by directly amplifying the spin-orbit coupling through the substitution of a heavy 5d element—as shown by introducing Re into Co$_2$CrGa—we achieved a colossal magnetostriction of -1008~ppm. This enhancement is driven by the synergistic effect of a larger SOC strength and a favorable modification of the electronic states near the Fermi level.

Finally, our comprehensive study has uncovered a general predictive framework for magnetostriction in these materials. We established a clear correlation between the magnetostriction coefficient and a descriptor combining the density of states at the Fermi level and the SOC strength of the Y-site atom. More importantly, we discovered a simple and powerful linear relationship between the magnetostriction and the choice of the Y-site transition metal for a fixed main-group element. These findings provide rational design principles that can guide future experimental efforts. This work not only broadens the potential applications of Heusler alloys into the domain of high-performance sensors and actuators but also paves the way for the accelerated discovery of new functional magnetic materials. We anticipate that these theoretical predictions will stimulate experimental validation of the most promising candidates identified herein.

\clearpage
\section{Acknowledgments}
This work is supported by the National Key R$\&$D program of China (Project Nos. 2022YFA1204000 and 2022YFA1402600) and the National Natural Science Foundation of China (Grant Nos. 12204346, 12274321, and 12361141823 ). 

\section*{Appendix}
\begin{table*}
	\centering
	\caption{Summary of first-principles results for the 25 Co-based full Heusler alloys. The table lists the calculated magnetoelastic coupling coefficient ($b_1$), the tetragonal shear modulus ($c'$), and the predicted tetragonal magnetostriction coefficient ($\lambda_{001}$). The unit for $b_1$ and $c'$ is eV per unit cell and the unit for $\lambda_{001}$ is parts per million (ppm).}
	\resizebox{0.45\textwidth}{!}{
		\begin{tabular}{m{1.7cm}m{1.7cm}m{1.7cm}m{1.7cm}} \hline\hline
			Co$_2$YZ & -b$_1$ &    c$'$ & $\lambda$$_{001}$ \\ \hline\hline
			Co$_2$VAl  & 12.8 & 118316.1 &  36  \\
			Co$_2$VGa & 15.4 & 89845 &    57   \\
			Co$_2$VSi & 30 & 95058 &    105  \\
			Co$_2$VGe  & 21 & 44675 &  156    \\
			Co$_2$VSn  & 28 & 82635 &  113  \\
			Co$_2$CrAl  & 27.66 & 81792.6 & 112  \\
			Co$_2$CrGa & 27 & 49411.9 & 184   \\
			Co$_2$CrSi & 16.1 & 80239.7 & 67   \\
			Co$_2$CrGe  & 11.5 & 71959.8 & 53   \\
			Co$_2$CrSn  & 23 & 83376 & 92  \\
			Co$_2$MnAl  & 23 & 104074.5 & 74   \\
			Co$_2$MnGa & 16.8 & 70605 & 79   \\
			Co$_2$MnSi & 7.8 & 125703.4 & 20  \\
			Co$_2$MnGe  & 10 & 100739 & 33   \\
			Co$_2$MnSn  & 9 & 102094 & 32  \\
			Co$_2$FeAl  & 5 & 109149 & 15   \\
			Co$_2$FeGa & 5 & 80581 & 20  \\
			Co$_2$FeSi & 13 & 119408 & 36   \\
			Co$_2$FeGe  & 11 & 84118 & 43    \\
			Co$_2$FeSn  & 12 & 87921 & 45  \\
			Co$_3$Al  & -27 & 51619 & -174   \\
			Co$_3$Ga & -26.3 & 36024 & -243  \\
			Co$_3$Si & -28.3 & 9761 & -966 \\
			Co$_3$Ge  & -22 & 27187 & -303  \\
			Co$_3$Sn  & -32.5 & 27815 & -385  \\  \hline\hline
			\label{tab2}		
	\end{tabular}}
\end{table*}

\begin{figure*}
	\includegraphics[width=18.5cm]{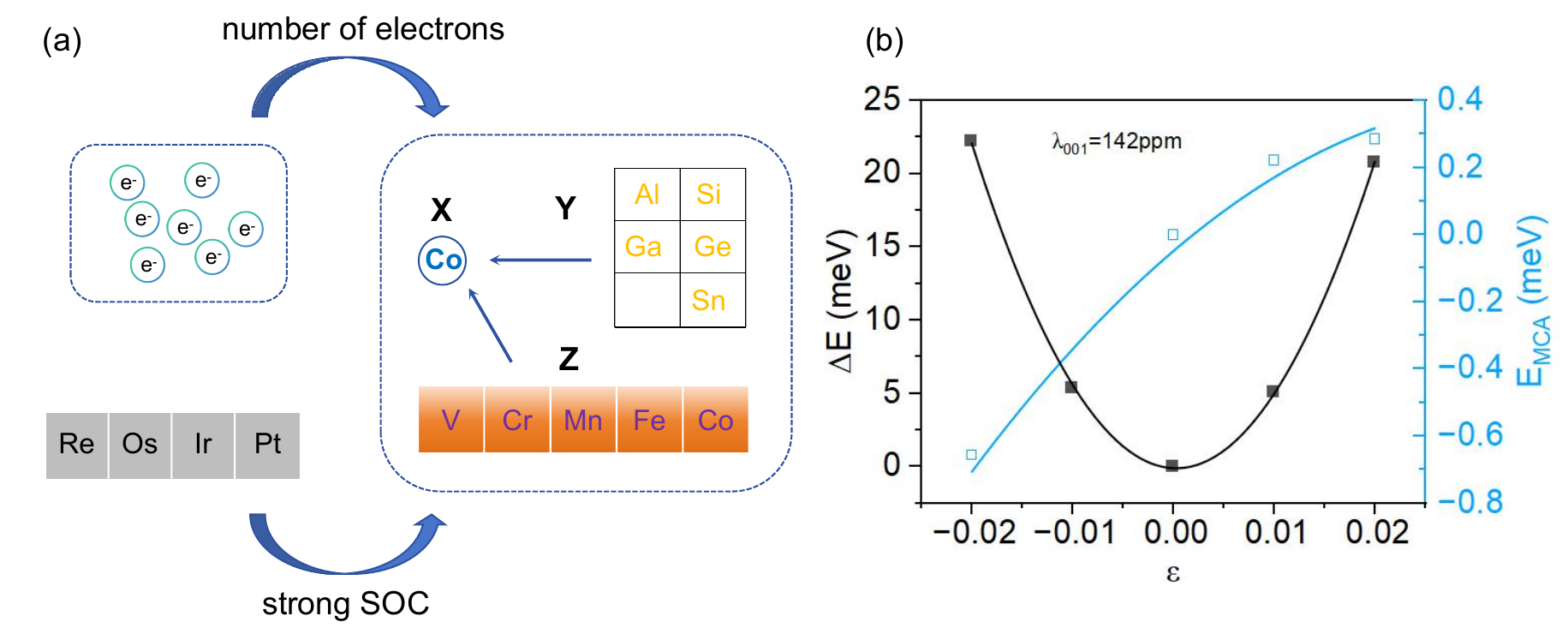}
	\caption{Illustration of the computational design and validation strategy. (a) the compositional factors for tuning magnetostriction in Co-based Heusler alloys, including the choice of Y-site transition metals (with varying SOC strength) and Z-site main group elements (which modulate the electron count). (b) the benchmark calculation performed on Fe$_7$Ga to verify the accuracy of our first-principles approach. Our calculated magnetostriction of 142 ppm for this system is in good agreement with established values.
		\label{fig5}}
\end{figure*}

\begin{figure*}
	\includegraphics[width=18.5cm]{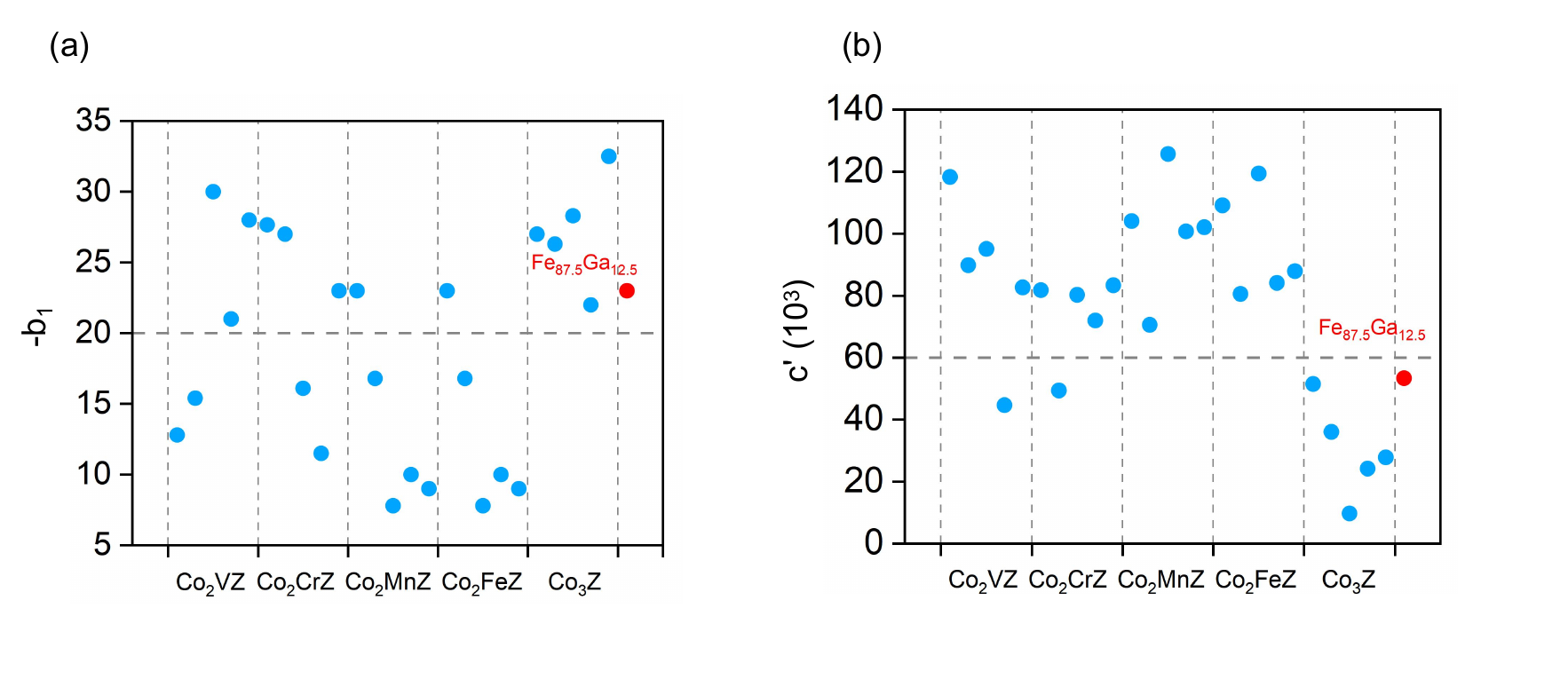}
	\caption{Comparison of (a) the calculated magnetoelastic coupling coefficient ($b_1$) and (b) the tetragonal shear modulus ($c'$) for the 25 Co-based Heusler alloys investigated in this work. The values for the reference compound Fe$_7$Ga are shown as horizontal dashed lines to provide a benchmark for materials with known large magnetostriction. These two parameters are the primary determinants of the magnetostriction coefficient $\lambda_{001}$.
		\label{fig6}}
\end{figure*}

\end{document}